\documentclass[12pt]{article}

\usepackage[english]{babel}

\usepackage[a4paper,top=2cm,bottom=2cm,left=3cm,right=3cm,marginparwidth=1.75cm]{geometry}

\usepackage{amsmath}
\usepackage{graphicx}
\usepackage[colorlinks=true, allcolors=blue]{hyperref}

\title{Flexible laboratory setup for DAC experimentation}
\author{Alfredo Pérez Vega-Leal, Manuel G. Satué}

\begin{document}
\maketitle

\begin{abstract}
Analog multiplexing appears to be a promising solution for modern transmitters, where speed is the primary limitation. The objective is the development of a low-cost solution to  compare different digital to analog (DAC) schemes. In particular, analog multiplexing techniques, high-speed single-DAC, Sigma-delta modulation, Dynamic element matching are considered. The work presents a review of these techniques and shows a  prototype of a time interleaved sigma delta modulation based DAC based on a commercially available Field Programmable Gate Array system.
\end{abstract}

\section{Introduction}

Digital to analog conversion is a process of transforming a digital signal into its corresponding analog signal. Digital signals having discrete amplitudes at discrete time instants are converted into a continuous signal both in case of amplitude and time in this process.

Digital to analog converters are widely used in countless applications at present. Almost all electronic devices are equipped with DACs. Due to its widespread use in almost all applications, there have been many researches to optimize the architecture of DAC for different applications \cite{su1994oversampling}.

Sigma-delta modulators (SDM) are known for their capability for encoding an analog or a multibit digital input signal into a two- or multilevel signal while preserving a high signal-to-noise ratio (SNR) \cite{kozak2003oversampled}. While the use of a two-level quantizer improves linearity of a DAC, it also adds a significant level of quantization noise. The  oversampling ratio (OSR) plays here an important role on SNR limitation \cite{colodro2020open,babaie2017mostly,chan2023high}.

Many proposals exist in the literature to cope with various factors affecting SDM-based DACs. This preprint reviews some of these factors as a background for the development of an unexpensive complete SDM-based DAC that can be used for experimentation.

\section{DAC Fundamentals and Limitations}
Currently, DACs are ubiquitous components utilized in a vast array of applications, with nearly all modern electronic devices incorporating them. Given this widespread adoption, extensive research has been conducted to optimize DAC architectures for specific operational requirements \cite{clara2012high}.

The precision of a DAC is fundamentally constrained by static errors in the output current or voltage levels, a phenomenon known as element mismatch. These
errors establish a ceiling for the achievable accuracy and resolution in high-precision applications, as they manifest as harmonic distortion, gain errors, and offset errors \cite{black1999analog}.

To counteract these non-idealities, various linearization methods are employed. These techniques can be summarized as physical level calibration, dynamic element matching (DEM), noise-shaping combined with digital calibration, and the application of large periodic high-frequency dithering or large stochastic high-pass noise dithering.

Noise and distortion in DACs arise from several sources: repeated
spectra, quantization, element mismatch, thermal noise, and semiconductor noise.
Specifically, element mismatch forces the DAC output levels to deviate from their ideal values; this results in a static error defined as Integral Non-Linearity (INL), which serves as a primary source of harmonic and intermodulation distortion \cite{barrett2022analog}.

In the domain of power electronics, discontinuous synchronized PWM strategies are extensively employed to ensure continuous phase voltage synchronization in voltage source inverters, a critical requirement for the efficient operation of electric drives \cite{oleschuk2012multiphase,shui2020current}.

Furthermore, PWM is a fundamental technique for encoding analog signals into
binary streams. While traditional implementations often suffer from harmonic distortion \cite{arahal2016harmonic}, advanced three-level PWM schemes have been demonstrated to significantly mitigate these distortion artifacts. This improvement enhances signal fidelity, making the architecture suitable for a broad spectrum of signal processing applications.

\section{Sigma-Delta Architectures}
Sigma-Delta Modulators (SDM) have become the architecture of choice for implementing high-resolution Digital-to-Analog Converters (DACs) due to their inherent linearity and robust analog implementation \cite{aziz1996overview}. 

By employing oversampling and noise-shaping techniques, these converters effectively suppress quantization noise within
the signal band, making them highly attractive for modern communication standards in both transmitter and receiver designs \cite{kester2009oversampling,pandita2009oversampling}.

A fundamental trade-off exists: moderate oversampling ratios (OSR) limit the achievable dynamic range \cite{orcioni2013dynamic,chang2010wide}, whereas high OSRs necessary for high resolution significantly restrict the available signal bandwidth \cite{colodro2010spectral,kozak2003oversampled,candy1992oversampling}.

To address these bandwidth limitations without compromising resolution, Multirate and Time-Interleaved (TI) architectures have been developed \cite{jiang2017reconfigurable,erfani2023low}. In multirate structures, the traditional multibit quantizer in the forward path can be replaced by a single-bit quantizer operating at a higher frequency \cite{cherry2002continuous,noh2025low}. This is achieved by increasing the oversampling ratio of the loop’s final integrators \cite{colodro2003multirate,jiang2017reconfigurable}. Further
architectural improvements in Time-Interleaved SDMs include replacing high-rate integrators with parallel low-rate units \cite{nguyen2022exploring}. This approach not only simplifies the design but also avoids the complications of delayed cross-paths \cite{yu2025high,petraglia2009switched,colodro2005time}.

Additionally, hardware complexity can be reduced by minimizing the number of comparator units through the use of a quantizer tracking loop clocked at a high frequency \cite{choi2021linearity,lan20212,colodro2010continuous}. Beyond standard linear conversion, specific challenges arise when implementing
quantizers based on Voltage-Controlled Oscillators (VCOs), which exhibit nonlinear voltage-frequency characteristics \cite{dehbovid2018nonlinear,chen2023effective}. To mitigate this, linearity enhancement techniques have been proposed where the signal is coarsely estimated from the VCO output and subtracted from the input \cite{borgmans2019enhanced}. This ensures the VCO operates solely on a small-signal component, thereby maintaining linearity \cite{choi2021linearity,han2019cmos,colodro2014linearity}.

Finally, in the realm of Field Programmable Gate Arrays (FPGAs) and analog
integrated circuits, data converters are essential for bridging the digital and analog domains \cite{uchagaonkar2012fpga,faghani2015comparison,todorovic2017implementation}. Analog IC technology continues to provide reliable, flexible solutions for complex systems \cite{kielbik2018aruz,monmasson2021system,colodro1996cellular}.

The implementation of oversampling DACs predominantly relies on Sigma-Delta
Modulators (SDM) \cite{geerts2002design}. Distinct from Nyquist-rate architectures, SDMs offer superior
linearity and resolution; however, the incorporation of non-linear internal quantizers
complicates stability analysis \cite{arahal2010stability}. This is relevant in regards to the determination of the stable  input amplitude range. This range typically contracts as the modulator order increases, a constraint that has prompted the development of numerous stabilization techniques for high-order topologies.

To transcend the bandwidth limitations inherent to standard SDMs, advanced
architectures such as Multirate and Time-Interleaving (TI) have been proposed \cite{caldwell2004time,talebzadeh2018novel}. Multirate strategies allow for the substitution of multibit quantizers with highspeed single-bit counterparts by manipulating the oversampling ratio of the final integration stages \cite{schmidt2020interleaving}.

Parallelization through Time-Interleaving further enhances throughput by distributing
the processing load \cite{dong2024low}. Notable implementations in this domain include
the use of parallel low-rate integrators to mitigate cross-path delays, and
the optimization of hardware resources by reducing comparator counts via fast-tracking
quantization loops \cite{colodro2022time,ghoneim2025design}. More recent innovations have introduced
analog-domain multiplexing to bolster robustness against path mismatches \cite{liu2025specialized}.

The versatility of Sigma-Delta modulation principles extends beyond traditional
digital-to-analog conversion, finding critical applications in specialized analog processing tasks. In the design of Phase-Locked Loops (PLLs), for instance, noiseshaping techniques have been successfully adapted to perform frequency-to-digital conversion \cite{zhao2022mem}. By sampling the Voltage Controlled Oscillator (VCO) output prior to feedback, the sampling error can be shaped via a third-order transfer function, analogous to the quantization error shaping in SDMs \cite{colodro2011frequency}.

\section{Dynamic Element Matching}

Despite the advantages of oversampling, multi-bit Sigma-Delta data converters face significant performance limitations due to static component mismatches. To address this, Dynamic Element Matching (DEM) has emerged as a standard solution.

DEM based DACs have been used  in oversampling
delta-sigma data converters, pipelined ADCs, and highresolution
Nyquist-rate DACs \cite{chan2008dynamic}. They have been proven to eliminate the negative effects of  component mismatches.  Without DEM, those mismatches cause nonlinear distortion. The DEM scrambles the pattern of use of the componentes, as a result, the error appears as pseudorandom noise,  uncorrelated
with the input sequence.

The efficacy of DEM relies on hardware redundancy. By employing multiple unit
elements, the system can generate a specific output level using various combinations of these elements. DEM algorithms leverage this redundancy to randomize or cyclically select the elements used for each sample. This process modulates the static mismatch error, transforming it into wideband noise that is uncorrelated with the signal and can be subsequently filtered out, thereby preserving the linearity of the DAC \cite{hamoui2004high}.

In this context, the primary objective of this work is to present the development of a cost-effective, comprehensive SDM-based DAC platform. This system is specifically engineered to facilitate the experimental validation and comparative analysis of diverse Dynamic Element Matching strategies \cite{colodro2009analog}. In particular, the data weighted averaging (DWA) method where the average number of times any DAC is used is equal to the rest. In this way the  errors produced by the element
mismatch can be spectrally shaped by a first-order high-pass transfer function  \cite{baird1995improved,chen2022dynamic}.

\section{Background material}

\subsection{Sigma delta modulator}
In Z domain, the output of an ADC can be expressed as 

$$Y(z) = X(z)H_x(z) + E(z)H_e(z)$$

\noindent where $H_x$ represents the transfer function of the signal and $H_e$ represents the transfer function of the noise. In many applications, $H_x$ is the unity transfer function, whereas $H_e$ can shape the frequency distribution of the noise to attain a higher output resolution. This is a characteristic of most SDM, where noise shaping attenuates the noise level in the signal band.

A first order SDM DAC uses an analog SDM followed by a digital decimator \cite{aziz1996overview}. From the signal $x[n]$, another signal is obtained by substracting an analog representation $y_a[n]$ of the quantized output $y[n]$, the difference is then filtered using $z^{-1}/(1-z^{-1})$. 

For an ideal DAC, the transfer function is unity. Then the output of the modulator is 

$$Y(z) = X(z)z^{-1} +  E(z) (1-z^{-1})$$

\noindent here, $H_e(z) = 1-z^{-1}$ contains a zero in $z=1$ causing infinite attenuation at DC, then large attenuation for lower frequencies and amplification at higher frequencies. The in-band noise power can be found as

$$\sigma_{ey}^2 = \sigma_e^2 \frac{\pi^2}{3} \frac{8 f_b^3}{f_s^3}$$

\subsection{Dynamic matching element}
The DEM methods are  digital techniques that reduce analog mismatch errors without requiring any a priori knowledge of the component variations.

In a DAC, the discrete-time input sequence $x(n)$ has produce  an analog signal $v(t)$. 

A three-level DAC can issue output
values $-\Delta$, $0$ and $+\Delta$. In practice, the sampled output is subject to additive noise resulting from mismatches among  circuit elements that should be identical. The DAC can then be viewed as a transforming block introducing a constant gain, a constant offset, and an additive error term $eDAC(n)$. The error term depends nonlinearly on $x(n)$.

The gain and offset pose little problems, however, the nonlinear term is difficult to handle. The DEM approach is based on converting $eDAC(n)$ into  pseudorandom noise uncorrelated with the input sequence.

\section{Laboratory benchmark}

\begin{figure}
    \centering
    
\begin{tabular}{c}
\begin{tabular}{ccc}
    \includegraphics[width=0.4\linewidth]{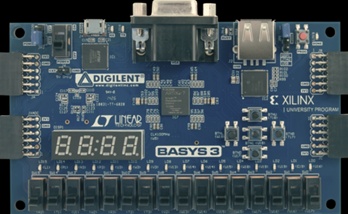} & $~$ &
    \includegraphics[width=0.2\linewidth]{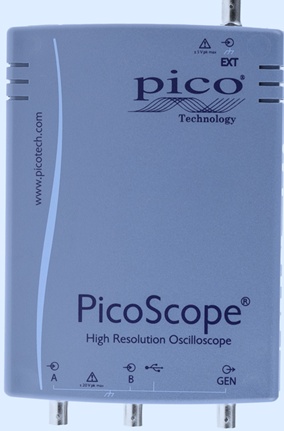} \\
        FPGA & & Oscilloscope \\
\end{tabular} \\    
    \includegraphics[width=0.75\linewidth]{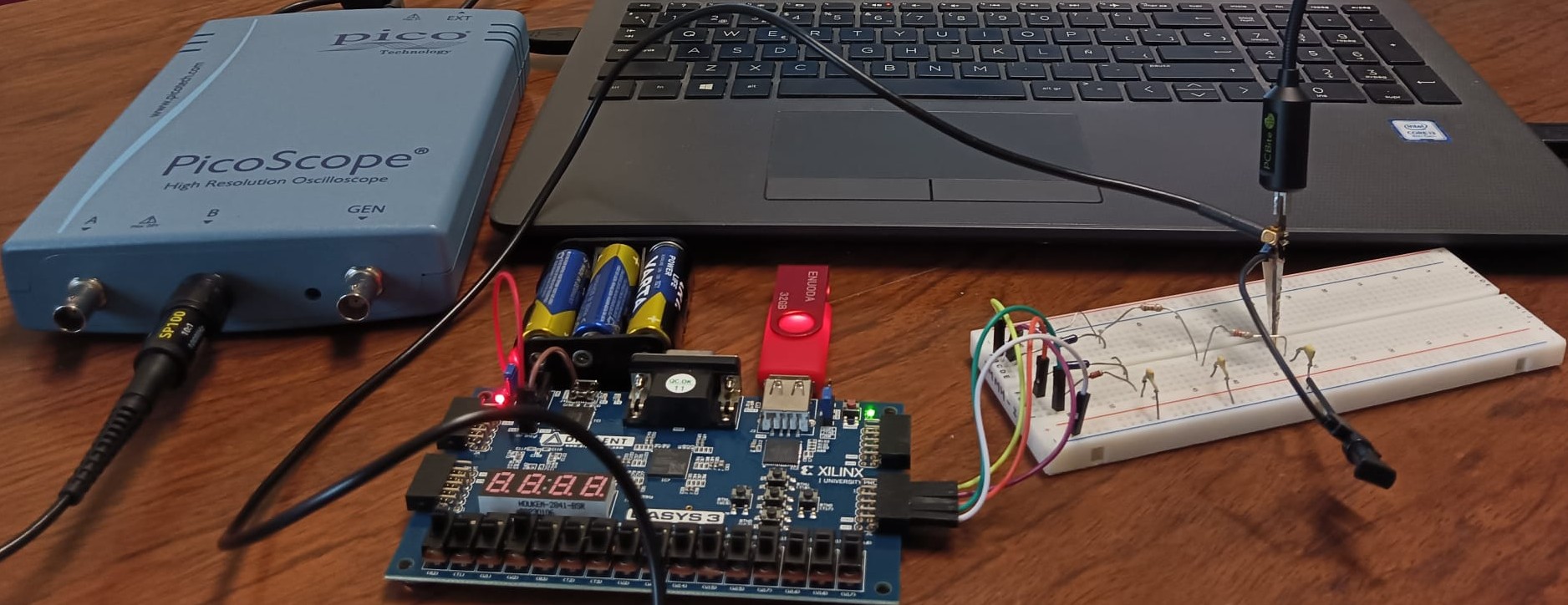} \\
    Overview \\
    \includegraphics[width=0.75\linewidth]{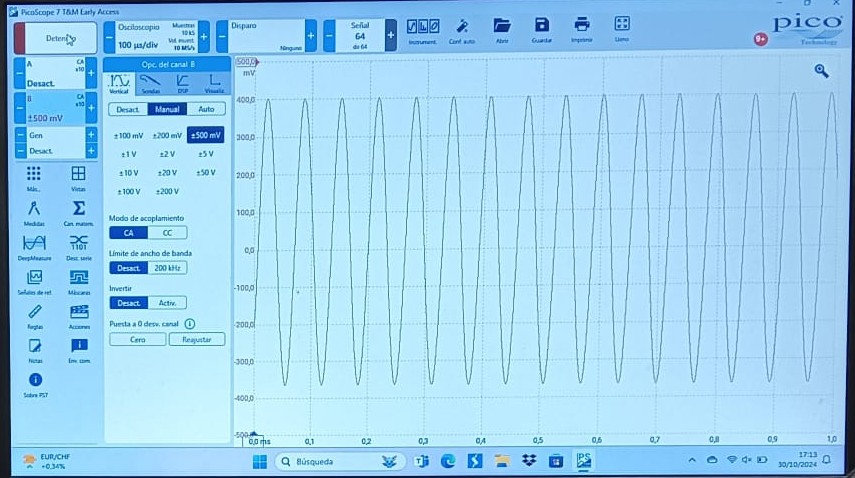} \\
    Oscilloscope output \\
\end{tabular}    
    \caption{Elements of the experimental setup.}
    \label{fig:exp-setup}
\end{figure}

A prototype of a TI SDM-based DAC has been implemented using low-cost easily available and flexible elements. The architecture for the proposed experimental setup is shown in Figure \ref{fig:exp-setup}.

In the prototype, digital logic operations are realized using programs that run on a Diligent BASYS-3 prototyping board. This element is a  digital circuit development platform  based on a 7 field FPGA  AMD Artix \textregistered XC7A35T-ICPG236C. It features low cost and a diverse connectivity for various applications. Its master clock runs at  $fck = 100$ MHz.

The needed sinusoidal signals are generated by the FPGA using a resonant circuit. Different clock frequencies $f_H$ can be obtained dividing the frequency of the master clock of the prototyping board; $f_H = f_{ck} / N$.

The outputs of the TI SDM, $yk(m)$, are the components that result from
the polyphase decomposition of $y(n)$.

\[y_k(m) = y(mM+k), k=0,1,..., M-1\]

Discrete-time frequency for  $y(n)$ is  $f_H$, wheras for signals $y_k(m)$ the sampling frequency is $fL$.

The inputs to the $M$ DACs can either originate from the $y_k(m)$ outputs of a TI SDM or from the
high-speed signal, $y(n)$, at the multiplexer output.

An FPGA output buffer has been used as the DAC in each path. The outputs of the DAC are added using a multiple-input single-output LPF. In turn, the output of the LPF can be monitored  using a digital oscilloscope.

The oscilloscope is a high-resolution Picoscope 4262. The main technical characteristics are: resolution 16 bits, maximum sampling frequency 10 Msps and a bandwidth of 5 MHz.

Initial tests have shown the versatility of the prototype to compare different techniques for research purposes.

\bibliographystyle{unsrt}
\bibliography{sample}

@inproceedings{baird1995improved,
  title={Improved Delta Sigma DAC linearity using data weighted averaging},
  author={Baird, Rex T and Fiez, Terri S},
  booktitle={Proceedings of ISCAS'95-International Symposium on Circuits and Systems},
  volume={1},
  pages={13--16},
  year={1995},
  organization={IEEE}
}

@article{orcioni2013dynamic,
  title={Dynamic OSR dithered sigma--delta modulation in solid state light dimming},
  author={Orcioni, Simone and d'Aparo, Rocco and Lobascio, Armando and Conti, Massimo},
  journal={International Journal of Circuit Theory and Applications},
  volume={41},
  number={4},
  pages={387--395},
  year={2013},
  publisher={Wiley Online Library}
}

@article{chang2010wide,
  title={Wide dynamic-range sigma--delta modulator with adaptive feed-forward coefficients},
  author={Chang, R-G and Chen, C-Y and Hong, J-H and Lee, S-Y},
  journal={IET circuits, devices \& systems},
  volume={4},
  number={2},
  pages={99--112},
  year={2010},
  publisher={IET}
}

@article{chen2022dynamic,
  title={A dynamic-range self-compensation technique in a noise shaping SAR ADC utilizing mismatch error shaping},
  author={Chen, Zhijie and Li, Hongyou and Jiang, Mengqian and Zhu, Sha and Wan, Peiyuan},
  journal={Electronics Letters},
  volume={58},
  number={10},
  pages={388--389},
  year={2022},
  publisher={Wiley Online Library}
}

@article{chan2008dynamic,
  title={Dynamic element matching to prevent nonlinear distortion from pulse-shape mismatches in high-resolution DACs},
  author={Chan, Kok Lim and Zhu, Jianyu and Galton, Ian},
  journal={IEEE Journal of Solid-State Circuits},
  volume={43},
  number={9},
  pages={2067--2078},
  year={2008},
  publisher={IEEE}
}

@article{hamoui2004high,
  title={High-order multibit modulators and pseudo data-weighted-averaging in low-oversampling Delta Sigma ADCs for broad-band applications},
  author={Hamoui, Anas A and Martin, Kenneth W},
  journal={IEEE Transactions on Circuits and Systems I: Regular Papers},
  volume={51},
  number={1},
  pages={72--85},
  year={2004},
  publisher={IEEE}
}

@article{arahal2016harmonic,
  title={Harmonic analysis of direct digital control of voltage inverters},
  author={Arahal, Manuel R and Barrero, Federico and Ortega, Manuel G and Martin, Cristina},
  journal={Mathematics and Computers in Simulation},
  volume={130},
  pages={155--166},
  year={2016},
  publisher={Elsevier}
}

@article{arahal2010stability,
  title={Stability analysis of five-phase induction motor drives with variable third harmonic injection},
  author={Arahal, MR and Duran, MJ and Barrero, F and Toral, SL},
  journal={Electric power systems research},
  volume={80},
  number={12},
  pages={1459--1468},
  year={2010},
  publisher={Elsevier}
}

@article{nguyen2022exploring,
  title={Exploring speed maximization of frequency-to-digital conversion for ultra-low-voltage VCO-based ADCs},
  author={Nguyen, Viet and Schembari, Filippo and Staszewski, Robert Bogdan},
  journal={IEEE Transactions on Circuits and Systems I: Regular Papers},
  volume={70},
  number={3},
  pages={1043--1056},
  year={2022},
  publisher={IEEE}
}

@article{noh2025low,
  title={A Low-Power, Wide-DR PPG Readout IC with VCO-Based Quantizer Embedded in Photodiode Driver Circuits},
  author={Noh, Haejun and Kim, Woojin and Kim, Yongkwon and Koh, Seok-Tae and Jeon, Hyuntak},
  journal={Electronics},
  volume={14},
  number={19},
  pages={3834},
  year={2025},
  publisher={MDPI}
}

@article{jiang2017reconfigurable,
  title={Reconfigurable mismatch-free time-interleaved bandpass sigma--delta modulator for wireless communications},
  author={Jiang, Dongyang and Sin, Sai-Weng and U, Seng-Pan and Martins, Rui Paulo and Maloberti, Franco},
  journal={Electronics Letters},
  volume={53},
  number={7},
  pages={506--508},
  year={2017},
  publisher={Wiley Online Library}
}

@book{cherry2002continuous,
  title={Continuous-time delta-sigma modulators for high-speed A/D conversion: theory, practice and fundamental performance limits},
  author={Cherry, James A and Snelgrove, W Martin},
  year={2002},
  publisher={Springer}
}

@article{erfani2023low,
  title={Low-clock-speed time-interleaved architecture for a polar delta--sigma modulator transmitter},
  author={Erfani Majd, Nasser and Fani, Rezvan},
  journal={ETRI Journal},
  volume={45},
  number={1},
  pages={150--162},
  year={2023},
  publisher={Wiley Online Library}
}

@incollection{zhao2022mem,
  title={An MEM Silicon Oscillating Accelerometer Employing a PLL and a Noise Shaping Frequency-to-Digital Converter},
  author={Zhao, Jian and Xu, Yong Ping and Su, Yan},
  booktitle={MEMS Silicon Oscillating Accelerometers and Readout Circuits},
  pages={93--131},
  year={2022},
  publisher={River Publishers}
}

@article{liu2025specialized,
  title={Specialized amplitude frequency error digital pre-equalizer for frequency interleaved DAC},
  author={Liu, Shengjian and Liu, Liansheng and Peng, Yu},
  journal={Measurement},
  pages={118914},
  year={2025},
  publisher={Elsevier}
}

@article{dong2024low,
  title={A low-power sigma-delta modulator based on high-order op-amp sharing technique for speech communication},
  author={Dong, Siwan and Ning, Sihao and Yuan, Menghan and Wang, Peng and Yang, Wenju},
  journal={AEU-International Journal of Electronics and Communications},
  volume={176},
  pages={155116},
  year={2024},
  publisher={Elsevier}
}

@inproceedings{ghoneim2025design,
  title={A Design Study of Power-Efficient Opamp Topologies for Discrete-Time Delta Sigma  Modulators},
  author={Ghoneim, Mahmoud and Ismail, Omar and Kauffman, John G and Ortmanns, Maurits},
  booktitle={2025 IEEE International Symposium on Circuits and Systems (ISCAS)},
  pages={1--5},
  year={2025},
  organization={IEEE}
}

@book{geerts2002design,
  title={Design of multi-bit delta-sigma A/D converters},
  author={Geerts, Yves and Steyaert, Michiel and Sansen, Willy},
  year={2002},
  publisher={Springer}
}

@book{caldwell2004time,
  title={Time-interleaved continuous-time delta-sigma modulators},
  author={Caldwell, Trevor C},
  year={2004}
}

@article{talebzadeh2018novel,
  title={A novel two-channel continuous-time time-interleaved 3rd-order sigma-delta modulator with integrator-sharing topology},
  author={Talebzadeh, Jafar and Kale, Izzet},
  journal={Analog Integrated Circuits and Signal Processing},
  volume={95},
  number={3},
  pages={375--385},
  year={2018},
  publisher={Springer}
}

@inproceedings{shui2020current,
  title={Current Harmonic Optimization of Two-Level Synchronous Symmetric SVPWM at Low Switching Frequency},
  author={Shui, Tianqing},
  booktitle={2020 Asia Energy and Electrical Engineering Symposium (AEEES)},
  pages={487--491},
  year={2020},
  organization={IEEE}
}

@article{pandita2009oversampling,
  title={Oversampling A/D converters with reduced sensitivity to DAC nonlinearities},
  author={Pandita, Bupesh and Martin, Kenneth W},
  journal={IEEE Transactions on Circuits and Systems II: Express Briefs},
  volume={56},
  number={11},
  pages={840--844},
  year={2009},
  publisher={IEEE}
}

@article{kester2009oversampling,
  title={Oversampling interpolating dacs},
  author={Kester, Walt},
  journal={MT-017, tutorial documents, Analog Devices, Inc. available at http://www. analog. com/static/imported-files/tutorials/MT-017. pdf},
  year={2009}
}

@article{kielbik2018aruz,
  title={ARUZ—Large-scale, massively parallel FPGA-based analyzer of real complex systems},
  author={Kie{\l}bik, Rafa{\l} and Ha{\l}agan, Krzysztof and Zatorski, Witold and Jung, Jaros{\l}aw and Ula{\'n}ski, Jacek and Napieralski, Andrzej and Rudnicki, Kamil and Amrozik, Piotr and Jab{\l}o{\'n}ski, Grzegorz and Sto{\.z}ek, Dominik and others},
  journal={Computer Physics Communications},
  volume={232},
  pages={22--34},
  year={2018},
  publisher={Elsevier}
}

@article{monmasson2021system,
  title={System-on-chip FPGA devices for complex electrical energy systems control},
  author={Monmasson, Eric and Hilairet, Micka{\"e}l and Spagnuolo, Giovanni and Cirstea, Marcian N},
  journal={IEEE Industrial Electronics Magazine},
  volume={16},
  number={2},
  pages={53--64},
  year={2021},
  publisher={IEEE}
}

@article{babaie2017mostly,
  title={A mostly digital VCO-based CT-SDM with third-order noise shaping},
  author={Babaie-Fishani, Amir and Rombouts, Pieter},
  journal={IEEE Journal of Solid-State Circuits},
  volume={52},
  number={8},
  pages={2141--2153},
  year={2017},
  publisher={IEEE}
}

@incollection{chan2023high,
  title={High-Performance Oversampling ADCs},
  author={Chan, Chi-Hang and Zhu, Yan and Qi, Liang and Sin, Sai Weng and Ortmanns, Maurits and Martins, Rui P},
  booktitle={Analog and Mixed-Signal Circuits in Nanoscale CMOS},
  pages={181--218},
  year={2023},
  publisher={Springer}
}

@article{candy1992oversampling,
  title={Oversampling methods for A/D and D/A conversion},
  author={Candy, James C and Temes, GC},
  journal={Oversampling delta-sigma data converters},
  pages={1--25},
  year={1992}
}

@article{black1999analog,
  title={Analog-to-digital converter architectures and choices for system design},
  author={Black, Brian},
  journal={Analog Dialogue},
  volume={33},
  number={8},
  pages={1--4},
  year={1999}
}

@incollection{barrett2022analog,
  title={Analog to digital conversion (adc)},
  author={Barrett, Steven F},
  booktitle={Arduino Microcontroller Processing for Everyone! Part II},
  pages={97--136},
  year={2022},
  publisher={Springer}
}

@book{schmidt2020interleaving,
  title={Interleaving concepts for digital-to-analog converters},
  author={Schmidt, Christian},
  year={2020},
  publisher={Springer}
}

@book{su1994oversampling,
  title={Oversampling digital-to-analog conversion},
  author={Su, David Kuochieh},
  year={1994},
  publisher={Stanford University}
}

@book{kozak2003oversampled,
  title={Oversampled delta-sigma modulators: Analysis, applications and novel topologies},
  author={Kozak, M{\"u}cahit and Kale, Izzet},
  year={2003},
  publisher={Springer}
}

@book{clara2012high,
  title={High-performance D/A-converters: Application to digital transceivers},
  author={Clara, Martin},
  volume={36},
  year={2012},
  publisher={Springer Science \& Business Media}
}

@article{choi2021linearity,
  title={Linearity enhancement of vco-based continuous-time delta-sigma adcs using digital feedback residue quantization},
  author={Choi, Moo-Yeol and Kong, Bai-Sun},
  journal={Electronics},
  volume={10},
  number={22},
  pages={2773},
  year={2021},
  publisher={MDPI}
}

@book{han2019cmos,
  title={CMOS Scaling Friendly Quantizers in Continuous-Time Delta-Sigma Modulators},
  author={Han, Changsok},
  year={2019},
  publisher={University of Florida}
}

@article{dehbovid2018nonlinear,
  title={Nonlinear analysis of VCO jitter generation using Volterra series},
  author={Dehbovid, Hadi and Adarang, Habib and Tavakoli, Mohammad Bagher},
  journal={COMPEL-The international journal for computation and mathematics in electrical and electronic engineering},
  volume={37},
  number={2},
  pages={755--771},
  year={2018},
  publisher={Emerald Publishing Limited}
}

@article{borgmans2019enhanced,
  title={Enhanced circuit for linear ring VCO-ADCs},
  author={Borgmans, Jonas and Rombouts, Pieter},
  journal={Electronics Letters},
  volume={55},
  number={10},
  pages={583--585},
  year={2019},
  publisher={Wiley Online Library}
}

@article{chen2023effective,
  title={An effective approach based on nonlinear spectrum and improved convolution neural network for analog circuit fault diagnosis},
  author={Chen, Le-rui and Khan, Umer Sadiq and Khattak, Muhammad Kashif and Wen, Sheng-jun and Wang, Hai-quan and Hu, He-yu},
  journal={Review of Scientific Instruments},
  volume={94},
  number={5},
  year={2023},
  publisher={AIP Publishing}
}

@inproceedings{todorovic2017implementation,
  title={Implementation and application of FPGA platform with digital MEMS microphone array},
  author={Todorovi{\'c}, Dejan and Salom, Iva and {\v{C}}elebi{\'c}, Vladimir and Prezelj, Jurij},
  booktitle={Proceedings of the Proceedings of 4th International Conference on Electrical, Electronics and Computing Engineering},
  pages={5--8},
  year={2017}
}

@inproceedings{faghani2015comparison,
  title={A comparison between two different FPGA-based topologies of first order sigma-delta modulator},
  author={Faghani, Maral and Isa, Maryam and Hamidon, Mohd Nizar and Mazaheri, Amin},
  booktitle={2015 IEEE International Circuits and Systems Symposium (ICSyS)},
  pages={120--124},
  year={2015},
  organization={IEEE}
}

@article{uchagaonkar2012fpga,
  title={FPGA based sigma--Delta analogue to digital converter design},
  author={Uchagaonkar, PA and Shinde, SA and Patil, VV and Kamat, RK},
  journal={International Journal of Electronics and Computer Science Engineering},
  volume={1},
  number={2},
  pages={508--513},
  year={2012}
}

@article{lan20212,
  title={A 2.5-GS/s Four-Way-Interleaved Ringamp-Based Pipelined-SAR ADC with Digital Background Calibration in 28-nm CMOS},
  author={Lan, Jingchao and Zhai, Danfeng and Chen, Yongzhen and Ni, Zhekan and Shen, Xingchen and Ye, Fan and Ren, Junyan},
  journal={Electronics},
  volume={10},
  number={24},
  pages={3173},
  year={2021},
  publisher={MDPI}
}

@article{colodro2010continuous,
  title={Continuous-time sigma--delta modulator with a fast tracking quantizer and reduced number of comparators},
  author={Colodro, F. and Torralba, A.},
  journal={IEEE Transactions on Circuits and Systems I: Regular Papers},
  volume={57},
  number={9},
  pages={2413--2425},
  year={2010},
  publisher={IEEE}
}

@inproceedings{colodro2005time,
  title={Time-interleaved multirate sigma-delta modulators},
  author={Colodro, F and Torralba, A. and Laguna, M},
  booktitle={2005 IEEE International Symposium on Circuits and Systems},
  pages={5581--5584},
  year={2005},
  organization={IEEE}
}

@article{yu2025high,
  title={A High Accuracy Input Stage of Current Sensing Isolated Amplifier},
  author={Yu, Wenxin and He, Lenian and Xi, Jianxiong},
  journal={Microelectronics Journal},
  pages={106932},
  year={2025},
  publisher={Elsevier}
}

@article{petraglia2009switched,
  title={Switched-capacitor decimation filter design using time-multiplexing and polyphase decomposition of transfer functions with low denominator orders},
  author={Petraglia, A. and Bar{\'u}qui, Fernando A. Pinto and Pereira, Jacqueline S},
  journal={Microelectronics journal},
  volume={40},
  number={12},
  pages={1673--1680},
  year={2009},
  publisher={Elsevier}
}

@article{colodro2003multirate,
  title={Multirate single-bit Sigma Delta modulators},
  author={Colodro, F. and Torralba, A.},
  journal={IEEE Transactions on Circuits and Systems II: Analog and Digital Signal Processing},
  volume={49},
  number={9},
  pages={629--634},
  year={2003},
  publisher={IEEE}
}

@inproceedings{colodro1996cellular,
  title={Cellular neuro-fuzzy networks {(CNFNs)}, a new class of cellular networks},
  author={Colodro, F and Torralba, A},
  booktitle={Proceedings of IEEE 5th International Fuzzy Systems},
  volume={1},
  pages={517--521},
  year={1996},
  organization={IEEE}
}

@inproceedings{oleschuk2012multiphase,
  title={Multiphase multi-inverter drive with discontinuous synchronized modulation},
  author={Oleschuk, Valentin and Gregor, R and Barrero, Federico},
  booktitle={2012 15th International Power Electronics and Motion Control Conference (EPE/PEMC)},
  pages={DS2a--8},
  year={2012},
  organization={IEEE}
}

@article{colodro2011frequency,
  title={Frequency-to-digital conversion based on sampled phase-locked loop with third-order noise shaping},
  author={Colodro, F and Torralba, A},
  journal={Electronics letters},
  volume={47},
  number={19},
  pages={1069--1070},
  year={2011},
  publisher={IET}
}

\end{document}